\documentclass[twocolumn,showpacs,preprintnumbers,amsmath,amssymb,fleqn]{revtex4} %% use 11pt for Applied Optics
\usepackage{amsmath,amssymb,graphicx}
\begin{document}

\title{Breaking mechanical dark mode via the Coulomb interaction}
\author{Jian-Song Zhang$^{1}$}
\author{Yuan Chen$^{1}$}
\author{Guang-Ling Cheng$^{1}$}
\author{Ai-Xi Chen$^{2}$}
\email{aixichen@zstu.edu.cn}
\affiliation{$^{1}$Department of Applied Physics, East China Jiaotong University,
Nanchang 330013, People's Republic of China\\
$^{2}$ Department of Physics, Zhejiang Sci-Tech University, Hangzhou 310018, People's Republic of China}

\begin{abstract}
We propose a method to break the dark mode of two degenerate mechanical resonators (MRs) in optomechanical systems via the Coulomb interaction.
Two degenerate MRs can be cooled to their ground-state simultaneously
beyond the resolved sideband regime
using the Coulomb interaction and an optical parametric
amplifier (OPA). We show that strong and robust mechanical squeezing
beyond 3 dB can be generated using the OPA and mechanical
parametric amplification (MPA) introduced by the Coulomb interaction.
Our results manifests that robust bipartite and genuine tripartite entanglement can be produced in a degenerate optomechanical system.
\end{abstract}
\maketitle

In recent years, many efforts have been devoted to the study of the ground-state cooling, strong squeezing,
and entanglement generation of MRs in optomechanical systems due to their wide applications such as highly sensitive measurement of tiny displacement, creation
of nonclassical states of mechanical modes, and quantum information processing \cite{Agarwal2013,Aspelmeyer2014,Bowen2015}.

Unfortunately, there are three main obstacles to realizing ground-state cooling, generating strong mechanical squeezing
and entanglement of two degenerate MRs in optomechanical systems.
The first obstacle is the so-called dark-mode effect of two degenerate MRs which suppresses the ground-state cooling of them
\cite{Genes2008,Dong2012,Wang2012,Shkarin2014,Kuzyk2017}. This effect has been demonstrated experimentally \cite{Ockeloen-Korppi2019}.
Until now, many methods have been employed to overcome the dark-mode effect \cite{Lai2018,Lai2020,Lai2021,Naseem2021,Sommer2019,Huang20221,Huang20222,Chu2024}.
For example, the dark mode can be broken by introducing a phase-dependent
phonon-exchange interaction \cite{Lai2020} or an auxiliary cavity mode \cite{Huang20221,Huang20222}.
One can also use the quantum reservoir engineering method \cite{Naseem2021} or the combination of the Duffing nonlinearities and OPA to
simultaneously cool degenerate MRs \cite{Chu2024}.
Very recently, the authors of Ref. \cite{Cao2025} have introduced a second cavity mode for the additional dissipative channel
to break the dark mode and cool two degenerate MRs simultaneously in experiments.
The second obstacle is the resolved sideband condition (the decay rate of an optical cavity must be much smaller than the frequencies of MRs),
which requires that the finesse of the optical cavity should be very high and limits the size of the
MRs to be cooled and entangled \cite{Agarwal2013,Aspelmeyer2014,Bowen2015}.
The third obstacle is the thermal fluctuations of the surrounding environment which will destroy the squeezing and entanglement of an optomechanical system.
Particularly, the bipartite and genuine tripartite entanglement of a quantum system are very fragile against thermal fluctuations.
To overcome these obstacles, we introduce the OPA and MPA induced by the Coulomb interaction \cite{Ma2014,Kong2017,Bai2017,Nejad2018,Mekonnen2023} in this work.

\par In this Letter, we propose a scheme to break the dark mode of two degenerate MRs
via the Coulomb interaction . As a result, two degenerate MRs can be cooled simultaneously beyond the resolved sideband regime with the help of
the Coulomb interaction and the OPA. Strong and robust mechanical squeezing is generated
even in the highly unresolved sideband regime.
In addition, robust bipartite and genuine tripartite entanglement
can be produced in this degenerate optomechanical system.

\par Here, we consider a system formed by two mechanical resonators that interact with a
cavity with frequency $\omega_a$ and decay rate $\kappa$. A degenerate OPA pumped by a laser field of frequency 2$\omega_d$ is
put into the cavity. We assume that the cavity is driven by a laser field of frequency $\omega_d$. The first MR is charged with positive charge
$Q_1 = |e| s \sigma_1 $, where $e$ is the electron charge, $s$ is the charged surface area,
and $\sigma_1$ is the surface charge density. The second MR is not charged.
Similar to Refs. \cite{Fengmang2013, Chenyuan2025}, there are two identical electrodes
which are distant by $2L$. Each electrode with the capacitance $C_0$ is controlled by the
bias gate voltage $U_0$. Here, $x_1$ and $x_2$ are the deviations of the first and second MRs from their equilibrium positions, respectively.
The total Hamiltonian of the present system is \cite{Fengmang2013, Chenyuan2025}
\begin{eqnarray}
H &=& H_0 + H_{dri} + H_I +H_{\chi} + H_{cou}, \\
H_0 &=& \hbar \omega_a a^{\dagger}a + \sum_{j=1,2} (\frac{p_j^2}{2m_j} + \frac{1}{2}m_j\omega_{m_j}^2 x_j^2),\\
H_{dri}&=& i\hbar E(e^{-i\omega_d t} a^{\dagger} - e^{i\omega_d t} a),\\
H_I&=& -\frac{\hbar \omega_a}{L} a^{\dagger}a(x_1 + x_2),\\
H_{\chi}&=&i\hbar \chi (e^{i\theta-2i\omega_d t}a^{\dagger 2}-e^{-i\theta+2i\omega_d t}a^{2}),\\
H_{cou}&=&\frac{C_0 U_0 Q_1}{4\pi\epsilon_0} (\frac{1}{L+x_1} + \frac{1}{L-x_1}),
\end{eqnarray}
with $E=\sqrt{\frac{\kappa \mathcal{P}}{\hbar \omega_d}}$, $\mathcal{P}$ being the power of the
driving field, $\epsilon_0$ being the vacuum dielectric constant. $H_0$ is the free Hamiltonian of
the whole system and $\omega_{m_j}$ is the frequency of the $j$th MR with position and momentum operator $x_j$ and $p_j$ satisfying the commutation relation $[x_j, p_j]=i\hbar$.
$H_{dri}$ denotes the Hamiltonian for driving field applied to the cavity mode.
$H_I$ describes the interactions between the cavity mode and two MRs.
The Hamiltonian of a second-order nonlinear medium is denoted
by $H_{\chi}$ and $\chi$ is the nonlinear gain of the OPA with
phase $\theta$. The last term $H_{cou}$ is the Hamiltonian related to the Coulomb interaction.
If the oscillation of the first MR is much smaller than the effective cavity length ($x/L \ll 1$),
then the Hamiltonian $H_{cou}$ can be recast as $H_{cou} \approx \frac{C_0 U_0 Q_1}{2\pi \epsilon_0 L} (1+\frac{x_1^2}{L^2})$ by truncating to the second order
term and neglecting the other higher order terms.
Similar to Refs. \cite{Fengmang2013, Chenyuan2025}, we get the total Hamiltonian in the frame rotating at the driving frequency $\omega_d$ as
\begin{eqnarray}
H &=& \hbar \Delta_a' a^{\dagger}a + \hbar \omega_{m_1}' b_1^{\dagger}b_1 + \hbar \omega_{m_2}b_2^{\dagger}b_2 + i\hbar E(a^{\dagger} - a) \nonumber \\
&&-\hbar \sum_{j=1,2} g_ja^{\dagger}a(b_j^{\dagger} + b_j) + i\hbar \chi (e^{i\theta}a^{\dagger 2} - e^{-i\theta}a^2) \nonumber\\
&&+ \hbar \lambda (b_1^{\dagger 2} + b_1^2), \label{H}
\end{eqnarray}
where $g_j = \omega_a x_{j, zpf}/L$, $x_j = x_{j,zpf}(b_j^{\dagger} + b_j)$,
$p_j=ip_{j,zpf}(b_j^{\dagger} - b_j)$, $x_{j, zpf} = \sqrt{\hbar/(2m_j\omega_{m_j})}$,
$p_{j, zpf} = \sqrt{\hbar m_j \omega_{m_j}/2}$,
$\Delta_a'=\omega_a - \omega_d$, $\omega_{m_1}'=\omega_{m_1} + 2\lambda$,
and $\lambda = C_0 U_0 Q_1/(4\pi\epsilon_0 L^3 m_1\omega_{m_1})$.
Note that the MPA is introduced by the Coulomb interaction as one can see from
the last term $\lambda (b_1^{\dagger 2} + b_1^2)$ of this equation.

The above Hamiltonian can be linearized using the following displacement transformations
$a \rightarrow \alpha + \delta a$ and $b_j \rightarrow \beta_j + \delta b_j$.
Using the Hamiltonian of Eq. (\ref{H}) and the displacement transformations, we obtain the following quantum Langevin equations
\begin{eqnarray}
\dot{\alpha} &=& -(i\Delta_a + \frac{\kappa}{2})\alpha + ig_1\alpha(\beta_1^* + \beta_1)\nonumber\\
             && +ig_2\alpha (\beta_2^* + \beta_2) + 2\chi e^{i\theta} \alpha^* + E, \nonumber\\
\dot{\beta}_1 &=& -(i\omega_{m_1}' + \frac{\gamma_1}{2})\beta_1 + ig_1|\alpha|^2 - 2i\lambda \beta_1^*,\nonumber\\
\dot{\beta}_2 &=& -(i\omega_{m_2} + \frac{\gamma_2}{2})\beta_2 + ig_2|\alpha|^2,\nonumber\\
\delta\dot{a} &=& -(i\Delta_a + \frac{\kappa}{2}) \delta a + i \widetilde{g}_1(\delta b_1^{\dagger} + \delta b_1) + i \widetilde{g}_2(\delta b_2^{\dagger} + \delta b_2) \nonumber\\
&&+ 2\chi e^{i\theta}\delta a^{\dagger} +\sqrt{\kappa} a^{in}(t), \label{QLEs} \\
\delta \dot{b}_1 &=& -(i\omega_{m_1}' + \frac{\gamma_1}{2}) \delta b_1 + i\widetilde{g}_1 (\delta a^{\dagger} + \delta a) \nonumber\\
&&-2i\lambda \delta b_1^{\dagger} + \sqrt{\gamma_1} b_1^{in}(t), \nonumber\\
\delta \dot{b}_2 &=& -(i\omega_{m_2} + \frac{\gamma_2}{2}) \delta b_2 + i\widetilde{g}_2 (\delta a^{\dagger} + \delta a) + \sqrt{\gamma_2} b_2^{in}(t), \nonumber
\end{eqnarray}
where $\gamma_j$ is the decay rate of the $j$th mechanical mode, $\widetilde{g}_j = g_j \alpha$,
$\Delta_a = \Delta_a' - g_1(\beta_1 + \beta_1^*) - g_2(\beta_2 + \beta_2^*)$.
If a proper phase of the input driving laser is chosen, then $\alpha$ is real.
Here, $a^{in}(t)$ is the zero-mean input vacuum noise of the cavity mode with correlation function $\langle a^{in}(t)a^{\dagger in}(t') \rangle = \delta(t-t')$;
$b_j^{in}(t)$ is the input mechanical thermal noise with correlation functions
$\langle b_j^{in}(t)b_j^{\dagger in}(t') \rangle = (n_{j, th} + 1)\delta(t-t')$
and $\langle b_j^{\dagger in}(t)b_j^{in}(t') \rangle = n_{j, th}\delta(t-t')$.
$n_{j,th} = 1/[e^{\hbar \omega_{m_j}/(k_B T_0)} - 1]$ is the initial thermal phonon number of the $j$th mechanical mode with $k_B$ and $T_0$ being the Boltzmann constant and the temperature of the heating bath, respectively.

Now, we define the following quadrature operators $X_{O = a, b_1, b_2} = (O^{\dagger} + O)/\sqrt{2}$
and $Y_{O = a, b_1, b_2} = i(O^{\dagger} - O)/\sqrt{2}$. The noise quadrature operators are defined as follows: $X_{O = a, b_1, b_2}^{in} = (O^{in \dagger} + O^{in})/\sqrt{2}$
and $Y_{O = a, b_1, b_2}^{in} = i(O^{in \dagger} - O^{in})/\sqrt{2}$.
From the above quantum Langevin equations, we get
\begin{eqnarray}
\dot{\vec{f}} = A \vec{f} + \vec{n}, \label{dfdt}
\end{eqnarray}
with $\vec{f} = (X_a, Y_a, X_{b_1}, Y_{b_1}, X_{b_2}, Y_{b_2})^T$,
$\vec{n} = (\sqrt{\kappa} X_a^{in}, \sqrt{\kappa} Y_a^{in}, \sqrt{\gamma_1} X_{b_1}^{in}, \sqrt{\gamma_1} Y_{b_1}^{in},\sqrt{\gamma_2} X_{b_2}^{in}, \sqrt{\gamma_2} Y_{b_2}^{in})^T$
and
\begin{eqnarray}
A = \left(
\begin{array}{cccccc}
\kappa_+ &  \Delta_{a,+}         & 0                 & 0 & 0 & 0 \\
\Delta_{a,-}  & \kappa_- &  2\widetilde{g}_1 & 0 & 2\widetilde{g}_2 & 0\\
0 & 0 & -\frac{\gamma_1}{2} & \omega_{m_1}& 0 & 0\\
2\widetilde{g}_1 & 0 & -\omega_{m_1} - 4\lambda  & -\frac{\gamma_1}{2} & 0 & 0\\
0& 0 &  0 & 0 & -\frac{\gamma_2}{2} & \omega_{m_2} \\
2\widetilde{g}_2 & 0 & 0  & 0 & -\omega_{m_2} & -\frac{\gamma_2}{2}
\end{array}
\right), \label{drift_matrix_A}
\end{eqnarray}
where $\kappa_{\pm} = -\frac{\kappa}{2} \pm 2\chi\cos\theta$ and $\Delta_{a,\pm} = \pm \Delta_a + 2\chi\sin\theta$.

Note that the dynamics of the present system can be described by a $6\times 6$ covariance matrix
$V$ with $V_{jk} = (\langle f_j f_k + f_kf_j\rangle)/2$. Using the definitions of the covariance matrix $V$, $\vec{f}$, and Eq. (\ref{dfdt}), we get the evolution of
the covariance matrix $V$ as follows:
\begin{eqnarray}
\dot{V} = A V + VA^T + D, \label{dVdt}
\end{eqnarray}
where $D$ denotes the noise correlation defined by $D = diag\{\kappa/2, \kappa/2, \gamma_1 (2n_{1,th} + 1)/2, \gamma_1 (2n_{1,th} + 1)/2,\gamma_2 (2n_{2,th} + 1)/2,
\gamma_2 (2n_{2,th} + 1)/2\}$.
In the following, we assume $\gamma_1 = \gamma_2$ and $n_{1, th} = n_{2, th} = n_{th}$ for simplicity.
According to the Routh-Hurwitz criterion \cite{DeJesus1987}, the system described by Eq. (\ref{dfdt}) is stable if and only if
all real parts of the eigenvalues of matrix A defined by Eq. (\ref{drift_matrix_A}) are
negative. All the parameters used in the present work satisfy the Routh-Hurwitz criterion, and the system is stable.

The steady-state mean phonon number $n_j$ of the $j$th mechanical mode is
$n_j = (V_{2j+1, 2j+1} + V_{2j+2, 2j+2}- 1)/2$.
Mechanical squeezing is defined as (in units of dB) \cite{Lv2015,Zhangrong2019,ZhangOE2020}
\begin{eqnarray}
S_{dB, b_j} = -10 \log_{10} (\langle \Delta X^2_{b_j}\rangle /\langle \Delta X^2\rangle_{zpf}) = -10 \log_{10} (2\langle \Delta X^2_{b_j} \rangle), \label{S_dB}
\end{eqnarray}
with $\langle \Delta X^2\rangle_{zpf} = 0.5$ being the zero-point fluctuations.
The bipartite entanglement between the cavity mode $a$ and the $j$th mechanical mode $b_j$ can be defined by logarithmic negativity \cite{Vidal2002}
\begin{eqnarray}
  E_N^{a|b_j} &=& \max[0,-\ln (2\tilde{v}_-^{a|b_j})], \label{E_N}
\end{eqnarray}
where $\tilde{v}_-^{a|b_j} = \min[eig(i \Omega P_{a|b_j} V_{a|b_j} P_{a|b_j})]$,
$\Omega = i\sigma_y \oplus i\sigma_y$, $P_{a|b_j} = \sigma_z \oplus$ 1,
$V_{a|b_j}$ is the covariance matrix of the cavity mode $a$ and the $j$th mechanical mode $b_j$.
Tripartite entanglement is quantified by the minimum residual contangle \cite{Adesso2006,Adesso2007,Qiu2022}
\begin{eqnarray}
R_{min} = \min[R_{\tau}^{a|b_1b_2},R_{\tau}^{b_1|ab_2},R_{\tau}^{b_2|ab_1}], \label{R_min}
\end{eqnarray}
with $R_{\tau}^{i|jk} = C_{i|jk}-C_{i|j}-C_{i|k} (i,j,k = a,b_1,b_2)$ being the residual contangle.
Here, $C_{u|v} = (E_N^{u|v})^2$, $u$ and $v$ can be one or two modes. When $v$ includes two modes,
$\Omega$ = $i\sigma_y \oplus i\sigma_y$ $\oplus i\sigma_y$,
$P_{a|b_j} V_{a|b_j} P_{a|b_j}$ is replaced by $P_{i|jk} V_{i|jk} P_{i|jk}$.
$P_{i|jk} = \sigma_z \oplus$ 1$\oplus$ 1, $V_{i|jk}$ denotes the covariance matrix of the whole system.
There is genuine tripartite entanglement in the system when the minimum residual entanglement $R_{min}$ is greater than zero.

Now, we discuss the influence of the OPA on ground-state cooling. In order to cool MRs efficiently, the Stokes heating process must
be significantly suppressed. Similar to Ref. \cite{Chenyuan2025}, we obtain a pair of optimal OPA
parameters
\begin{eqnarray}
\chi_{opt} &=& \frac{\sqrt{0.25 \kappa^2 + (\Delta_a - \omega_{m_1}')^2}}{2},\label{optimal_x}\\
\theta_{opt} &=& i \ln[\frac{i(\Delta_a - \omega_{m_1}') - 0.5\kappa}{2\chi_{opt}}], \label{optimal_theta}
\end{eqnarray}
with $\omega_{m_1}' = \omega_{m_1} + 2\lambda$. In this case, the heating process of MRs can be suppressed.
Physically, this effect is a natural result of the destructive interference between
the down-converted low frequency photon from the Stokes process and the squeezed intracavity photon created by the OPA
\cite{Asjad2019}. In particular, the anti-Stokes cooling rate is the same as that of the bare cavity case
and the resolved sideband restriction can be relaxed \cite{Chenyuan2025}.

In our numerical simulation, the parameters are in accordance with the
current experimental level \cite{Fengmang2013,Chenyuan2025,LaHaye2004,Thompson2008,Hensinger2005}:
$m=20$ pg, $\omega_{m_1} = \omega_{m_2} = \omega_m = 2\pi \times 134$ kHz,
$C_0 =27.5$ nF, $\gamma_1=\gamma_2 = 10^{-6}\omega_m$, $L=0.1mm$, $\sigma_1 = 1.25\times 10^{13} cm^{-2}$
and $s = 0.08 \mu m^2$.

\begin{figure}[tbp]
\centering {\scalebox{0.3}[0.3]{\includegraphics{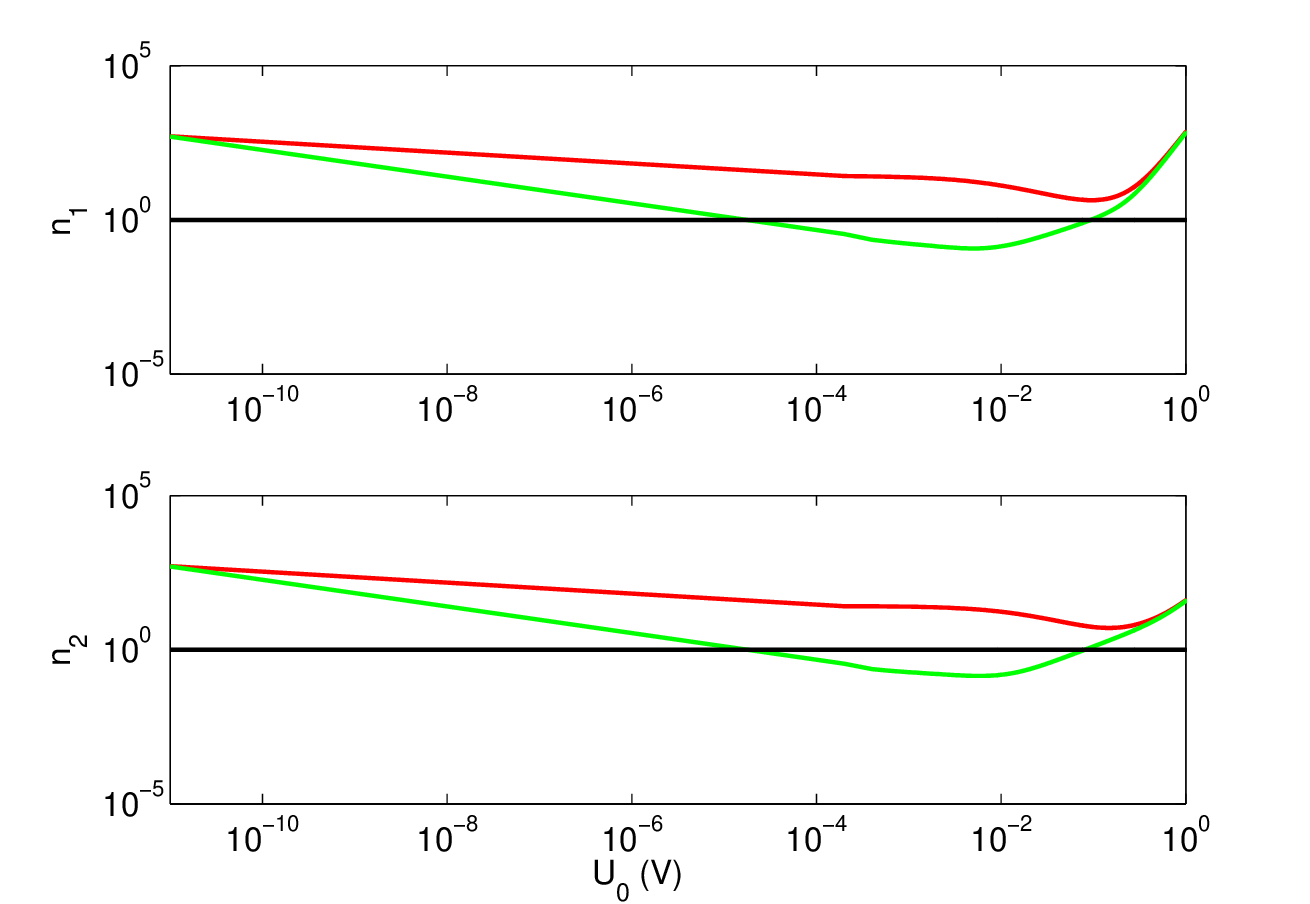}}}
\caption{Steady-state mean phonon numbers $n_1$ and $n_2$ versus the bias gate voltage $U_0$ for $\chi = 0$ (red line) and $\chi = \chi_{opt} = \kappa/4$ (green line) with
$\theta = \theta_{opt}= \pi$, $\Delta_a = \omega_m'$, $\widetilde{g}_1 = \widetilde{g}_2 = 0.3\omega_m$, $n_{th} = 1000$, and $\kappa = 20\omega_m$.
Other parameters are illustrated in the main text.}\label{fig1}
\end{figure}

\begin{figure}[tbp]
\centering {\scalebox{0.3}[0.3]{\includegraphics{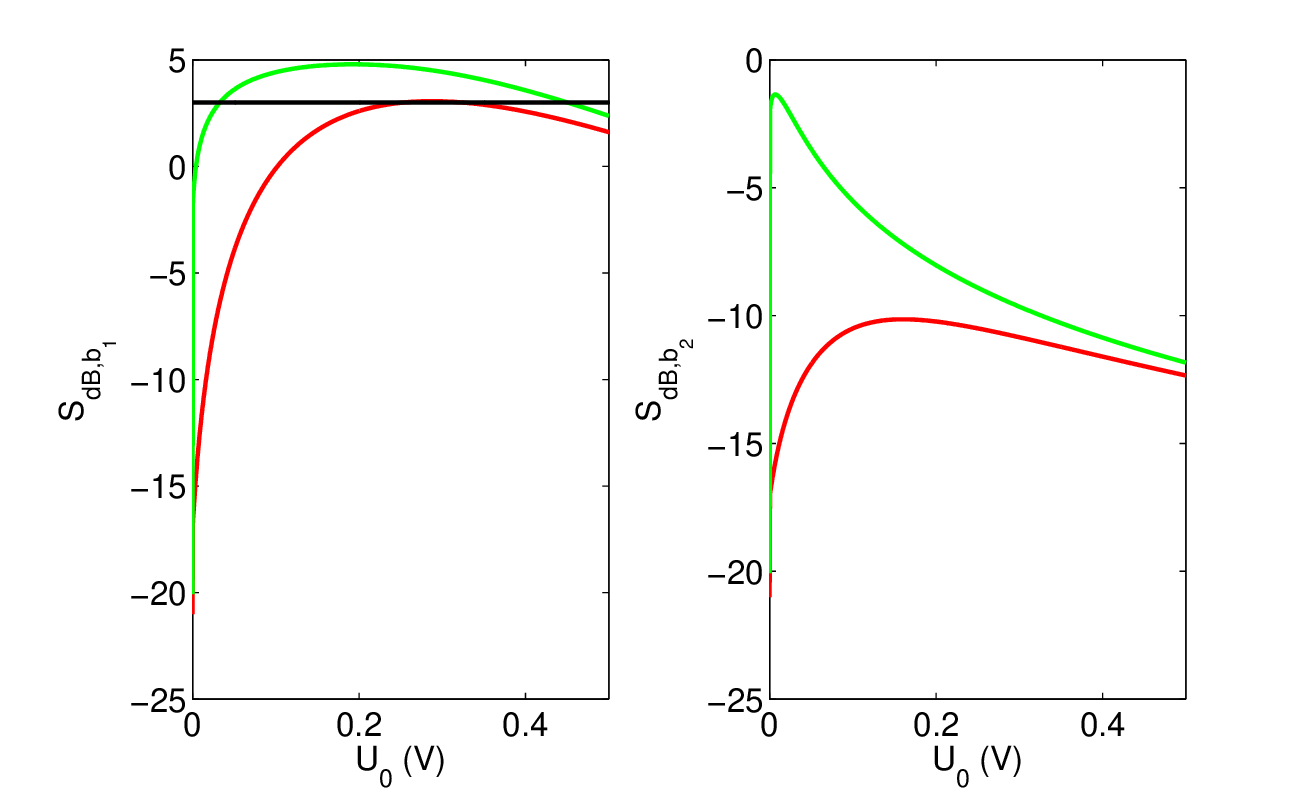} }}
\caption{Steady-state mechanical squeezing  versus the bias gate voltage $U_0$
for $\chi = 0$ (red line) and $\chi = \chi_{opt} = \kappa/4$ (green line) with
$\theta = \theta_{opt} = \pi$, $\Delta_a = \omega_m'$, $\widetilde{g}_1 = \widetilde{g}_2 = 0.3\omega_m$, $n_{th} = 100$, and $\kappa = 20\omega_m$.
The black line corresponds to mechanical squeezing at 3 dB.}\label{fig2}
\end{figure}

\begin{figure}[tbp]
\centering {\scalebox{0.3}[0.3]{\includegraphics{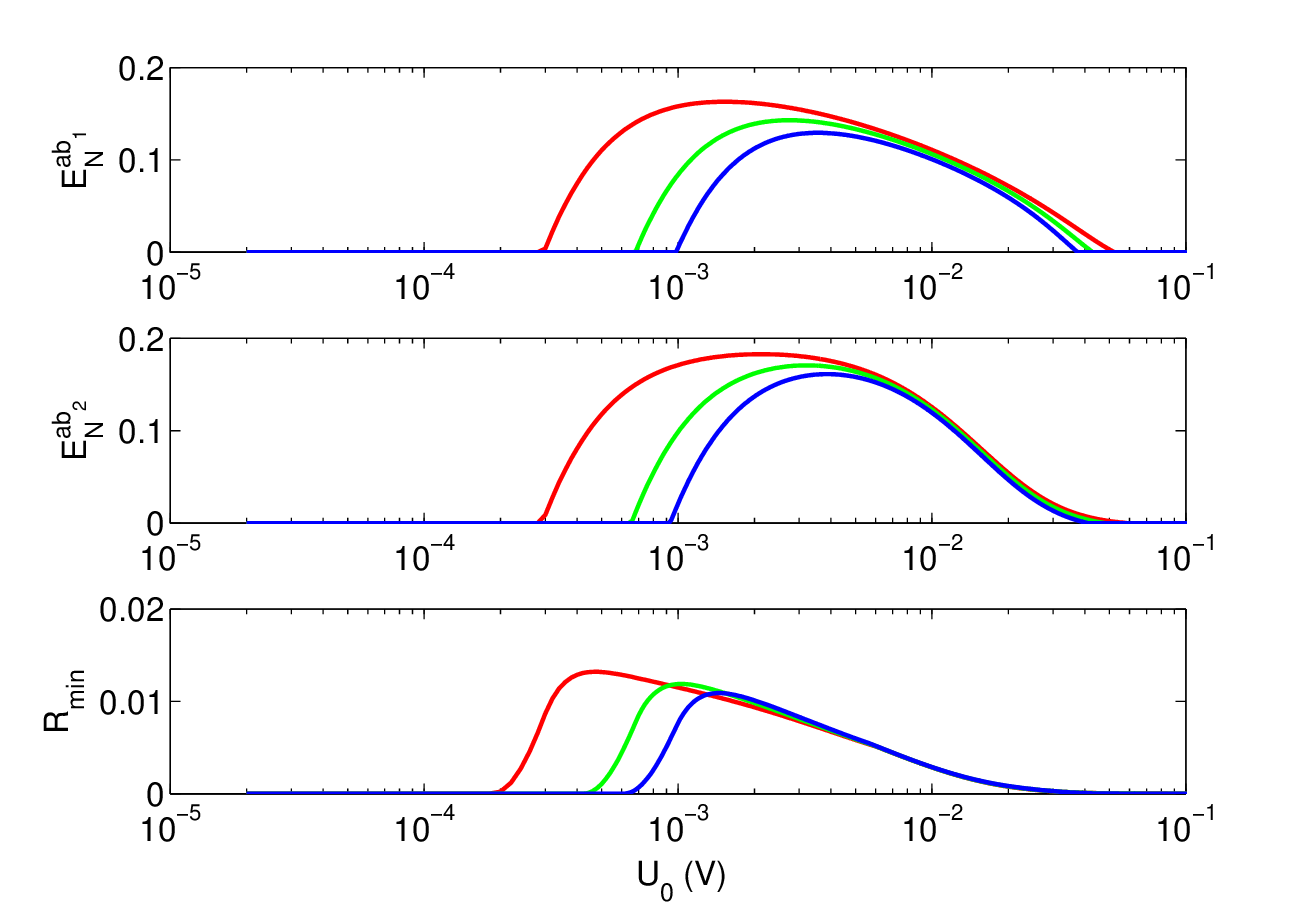}}}
\caption{Steady-state bipartite entanglement $E_N^{a|b_1}$, $E_N^{a|b_2}$, and genuine tripartite entanglement $R_{min}$
versus the bias gate voltage $U_0$ for $n_{th} = 100$ (red line),
$n_{th} = 500$ (green line), and $n_{th} = 1000$ (blue line) with
$\chi = \chi_{opt} = \kappa/4$, $\theta = \theta_{opt}= \pi$, $\Delta_a = \omega_m'$, $\widetilde{g}_1 = \widetilde{g}_2 = 0.3\omega_m$, and $\kappa = \omega_m$.}\label{fig3}
\end{figure}

In Fig. 1, we plot the steady-state mean phonon numbers $n_1$ and $n_2$ as functions of the bias gate voltage $U_0$ for different
values of $\chi$ with $n_{th} = 1000$ and $\kappa = 20\omega_m$. In the absence of the bias gate voltage ($U_0=0$),
two degenerate MRs cannot be efficiently cooled since they can form a dark mode.
However, the dark mode formed by two degenerate MRs can be broken using the Coulomb interaction.
The reason is as follows. The effective frequency of the first mechanical mode $b_1$ is $\omega_{m_1}'=\omega_{m_1} + 2\lambda$
as one can see from Eq. (\ref{H}). If the two MRs are degenerate with $\omega_{m_1} = \omega_{m_2}$, the effective
frequency of the first mechanical mode is different from that of the second mechanical mode in the presence of the Coulomb interaction with
$\lambda = C_0 U_0 Q_1/(4\pi\epsilon_0 L^3 m_1\omega_{m_1})$.
This implies that the dark mode of two degenerate MRs is broken via the Coulomb interaction.
To explain this more clearly, we obtain the effective Hamiltonian of Eqs. (\ref{QLEs}) under the rotating wave approximation
\begin{eqnarray}
H_{eff} &=& \Delta_a \delta a^{\dagger} \delta a + \omega_B B^{\dagger} B + \omega_D D^{\dagger}D \nonumber\\
&& + [\widetilde{g}_+ \delta a^{\dagger} B + \omega_{B} ^{(\lambda)}(B^{\dagger 2} + B^{\dagger}B)+ \omega_{D} ^{(\lambda)}(D^{\dagger 2} + D^{\dagger}D)  \nonumber\\
&& + g_{BD}^{(\omega)} B^{\dagger}D + 2 g_{BD}^{(\lambda)} (B^{\dagger}D^{\dagger} + B^{\dagger}D) \nonumber\\
&& + i\chi e^{i\theta} \delta a^{\dagger 2} + H.c.], \label{H_eff}
\end{eqnarray}
with $B = (\widetilde{g}_1 \delta b_1 + \widetilde{g}_2 \delta b_2)/\widetilde{g}_+$,
$D = (\widetilde{g}_2 \delta b_1 - \widetilde{g}_1 \delta b_2)/\widetilde{g}_+$,
$\widetilde{g}_+ = \sqrt{\widetilde{g}_1^2 + \widetilde{g}_2^2}$,
$\omega_B = (\omega_{m_1}\widetilde{g}_1^2 + \omega_{m_2}\widetilde{g}_2^2)/\widetilde{g}_+^2$,
$\omega_D = (\omega_{m_1}\widetilde{g}_2^2 + \omega_{m_2}\widetilde{g}_1^2)/\widetilde{g}_+^2$,
$\omega_{B} ^{(\lambda)} = \lambda \widetilde{g}_1^2/\widetilde{g}_+^2$,
$\omega_{D} ^{(\lambda)} = \lambda \widetilde{g}_2^2/\widetilde{g}_+^2$,
$g_{BD}^{(\omega)} = (\omega_{m_1} - \omega_{m_2}) \widetilde{g}_1\widetilde{g}_2/\widetilde{g}_+^2$,
and $g_{BD}^{(\lambda)} = \lambda \widetilde{g}_1\widetilde{g}_2/\widetilde{g}_+^2$.
In the absence of the Coulomb interaction [$g_{BD}^{(\lambda)} = 0$], two degenerate MRs with
$g_{BD}^{(\omega)} = 0$ can form a dark mode $D$ since it is totally decoupled from $B$ and $\delta a$ simultaneously
in this case as one can clearly see from the third line of Eq. (\ref{H_eff}).
Thus, the dark mode cannot be cooled
and it is impossible to efficiently cool two degenerate MRs in the case of $g_{BD}^{(\lambda)} = 0$
and $g_{BD}^{(\omega)} = 0$.
Fortunately, the dark mode formed by two degenerate mechanical modes can be destroyed
by introducing the Coulomb interaction, i.e. $g_{BD}^{(\lambda)} = \lambda \widetilde{g}_1\widetilde{g}_2/\widetilde{g}_+^2 > 0$,
since there is a direct interaction between two modes $B$ and $D$.
The cavity mode interacts with the hybrid mode $D$ indirectly with the help of the mode $B$.
Consequently, the thermal excitations of $B$ and $D$ can be efficiently extracted through the cooling channel of the optical mode.
It is possible to realize simultaneous ground-state cooling
of degenerate mechanical modes with the help of the Coulomb interaction.
In addition, the Stokes heating process can be significantly suppressed when the OPA parameters are given by
Eqs. (\ref{optimal_x}) - (\ref{optimal_theta}). Thus, two degenerate MRs can
be cooled to their ground-state simultaneously with the help of the OPA and Coulomb interaction
in a highly unresolved sideband regime. From the green lines of Fig. 1, one can observe that
the steady-state mean phonon number $n_1$ ($n_2$) first decreases with the
bias gate voltage $U_0$ and reaches the minimal value 0.12 (0.14) at $U_0 \approx 0.006$ V ($U_0 \approx 0.007$ V), then increases with the bias gate voltage.
This can be explained as follows.
On the one hand, the dark mode is broken by the Coulomb interaction. This is helpful for ground-state cooling.
On the other hand, the effective thermal phonon number of the first MR increases with the strength of the Coulomb interaction \cite{Chenyuan2025} which
is harmful to ground-state cooling. Consequently, the minimal values of $n_1$ and $n_2$ are
tradeoffs between these two competing effects.

In Fig. 2, we show the steady-state mechanical squeezing $S_{dB, b_1}$ and $S_{dB, b_2}$ as functions of the bias gate voltage $U_0$.
In the absence of the OPA ($\chi = 0$), the steady-state mechanical squeezing is slightly greater than 3 dB for $0.26 V \lesssim  U_0 \lesssim  0.32 V$.
In most cases, $S_{dB, b_1}$ is less than 3 dB as one can see from the red line of this figure.
Note that a strong mechanical squeezing of the first MR greater than 3 dB is generated for $0.036 V \lesssim  U_0 \lesssim  0.45 V$
in the presence of the OPA with $\chi = \chi_{opt} = \kappa/4$ and $\theta = \theta_{opt} = \pi$.
We now discuss the influence of the Coulomb interaction on the mechanical squeezing of the first MR.
Firstly, the Coulomb interaction can introduce MPA as one can clearly see from the last term $\lambda (b_1^{\dagger 2} + b_1^2)$
of Eq. (\ref{H}) with $\lambda = C_0 U_0 Q_1/(4\pi\epsilon_0 L^3 m_1\omega_{m_1})$. This term can be used to generate strong mechanical squeezing
of the first MR.
The parameter $\lambda$ could be conveniently controlled by the bias gate voltage $U_0$.
Secondly, the dark mode can be broken via the Coulomb interaction.
In fact, if the bias gate voltage is zero, the dark mode of two MRs is generated
and the mechanical squeezing is completely destroyed. Thus, the Coulomb interaction plays a constructive role in the generation of mechanical squeezing.
However, the effective thermal phonon number of the first MR increases with the Coulomb interaction, as we have pointed out previously \cite{Chenyuan2025}.
In this context, the Coulomb interaction plays a destructive role in the generation of mechanical squeezing.
Consequently, there is an optimal bias gate voltage for the generation of mechanical squeezing.
Note that there is no Coulomb interaction and MPA of the second MR. The mechanical squeezing of the second MR
is destroyed by the thermal fluctuations.

In Fig. 3, we plot the steady-state bipartite and genuine tripartite entanglement as functions of the bias gate voltage $U_0$ for different values of
$n_{th}$. In general, entanglement is very fragile and is easily disrupted by thermal fluctuations.
From this figure, we observe that robust bipartite and genuine tripartite entanglement can be produced in the present model.
In the absence of the Coulomb interaction, there is no bipartite or tripartite entanglement since
they are totally destroyed by the dark mode. If the bias gate voltage is applied, the dark mode is broken, and the bipartite
and tripartite entanglement can be generated. If $U_0$ is large enough, $E_N^{a|b_1}$, $E_N^{a|b_2}$ and $R_{min}$ are zero since
the effective thermal phonon number of the first MR, which is harmful to the generation of entanglement,
increases with the strength of the Coulomb interaction.
The amount of entanglement in the present system decreases with the parameter $n_{th}$. However, there are
bipartite and genuine tripartite entanglement even when $n_{th} = 1000$, i.e., robust entanglement can be produced.

In summary, we have presented the combination of the Coulomb interaction and OPA to
break the dark mode of two degenerate MRs in the highly unresolved sideband regime.
The Coulomb interaction can be conveniently adjusted by the
bias gate voltage. The simultaneous ground-state cooling of two degenerate MRs can be achieved within the reach of current
technology. We have shown that robust and strong mechanical squeezing (larger than 3 dB) of the MR with the
Coulomb interaction can be produced. In particular, the steady-state bipartite and genuine tripartite entanglement
of the system can be generated which are robust against the thermal fluctuations of its environment.
Our work opens up a new way to achieve ground-state cooling, strong mechanical squeezing, robust bipartite and tripartite entanglement
in degenerate optomechanical systems in the highly unresolved sideband regime.

\section*{Funding.}
National Natural Science Foundation of China (12465002,12465003,12575031,12547108); Natural Science Foundation of Jiangxi Province (20232ACB201013)

\section*{Disclosures.} The authors declare no conflicts of interest.

\end{document}